\documentstyle[11pt,titlepage]{article}  
\titlepage  
 
\begin{document} 
\begin{center} 
{\Large \bf Canonical Gravity, Diffeomorphisms
and Objective Histories} 
\vskip 1cm 
{\large Joseph Samuel} \\ 
Raman Research Institute\\ 
Bangalore 560 080, INDIA.\\ 
\end{center} 
\vskip 2 cm 
\vskip 5 cm 
email:sam@rri.ernet.in 
\newpage  
\section*{Abstract} 
This paper discusses the implementation
of diffeomorphism invariance in purely Hamiltonian 
formulations of General Relativity.
We observe that, if a constrained Hamiltonian formulation
derives from a manifestly covariant Lagrangian, 
the diffeomorphism invariance of the Lagrangian
results in the following properties of the constrained 
Hamiltonian theory:
the diffeomorphisms are 
generated by constraints on the phase space so that
a) The algebra of the generators reflects the algebra of the diffeomorphism
group. b) The Poisson brackets of the basic fields 
with the generators reflects the space-time transformation properties
of these basic fields. This suggests that {\it in a purely Hamiltonian
approach} the requirement of diffeomorphism invariance should 
be interpreted to include b) and not 
just a) as one might naively suppose. Giving up b) amounts to
giving up objective histories, even at the classical level.
This observation has implications for Loop Quantum Gravity
which are spelled out in a companion paper. We also describe an 
analogy between canonical gravity and Relativistic particle
dynamics to illustrate our main point.
\newpage
\section{Introduction}
The diffeomorphism invariance of General Relativity presents
both conceptual and technical problems\cite{general} for
quantisation. At the conceptual level, it leads to
deep questions about the nature of time, observables and the 
interpretation of quantum theory.
At the technical level, diffeomorphism invariance leads to
constraints on the classical phase space \cite{general}, 
which in a quantum theory, must
be imposed on physical states. Solving these
constraints has occupied much of the effort in the canonical
approach to quantum gravity. Several constrained 
Hamiltonian formulations (CHFs) of General Relativity 
exist today, each with its own following. 
It remains to be seen which of these formulations
will be the most advantageous in the approach to
quantum General Relativity.

This paper seeks to clarify the meaning of 
diffeomorphism invariance in a classical, constrained 
Hamiltonian Theory. Given a constrained theory, how does 
one test for diffeomorphism invariance?
The answer to this question involves a  subtlety, on which we 
focus in this paper. There is a substantial literature \cite{Tei,Ku1,Ku2,Ho,
Ku3,Po,Cla}
on the constrained Hamiltonian formulation of diffeomorphism
invariant theories. The point we wish to emphasise here 
is perhaps implicit in these earlier works, but we wish to
make it explicit in order to use it in \cite{ham}.
 Our strategy in addressing this question will be 
 to start with CHF's which we {\it know} are diffeomorphism
 invariant: those that are derived by a Legendre transformation
 from a manifestly covariant
 Lagrangian. We will then notice that 
the resulting constrained Hamiltonian
formulation satisfies certain conditions as a consequence
of the diffeomorphism invariance of the Lagrangian starting
point. We will explicitly spell out these conditions and
use these as a criterion for testing for 
diffeomorphism invariance
even when a Lagrangian starting point
is not available.
For example many currently popular CHFs of General Relativity 
\cite{Ash,Barbero} are
derived by making canonical transformations on the phase space; they
are entirely Hamiltonian in spirit and are often
presented and discussed without any Lagrangian starting point.
One would like to discuss the 
diffeomorphism invariance of such formulations in
a purely Hamiltonian framework.
The purpose of this paper is
to clarify how this can be done.

 The paper is organised as follows: In section II,
we recapitulate  some known results about the gauge
symmetries of Lagrangian systems and show how these
symmetries manifest themselves in a Hamiltonian framework.
In section III we illustrate these general results
using familiar examples like the ADM formalism,
gravity in 2+1 dimensions and Ashtekar's extended phase space
construction (EPS). 
In section IV,
we distinguish between strong and weak 
diffeomorphism invariance of a CHF
and 
bring out an analogy 
with a much simpler situation:
relativistic particle dynamics. 
Section V is a concluding discussion.

\section{Symmetries of Singular Lagrangian systems}
Consider a dynamical system with configuration manifold 
${\cal Q}$ on which local co-ordinates are $q^r,r=1..n$. The tangent
bundle over ${\cal Q}$ is $T{\cal Q}$ and the Lagrangian $L(q,{\dot q})$ 
is a real valued
function on $T{\cal Q}$. The Lagrangian $L$ defines a map from 
$T{\cal Q}$ to the cotangent bundle $T^*{\cal Q}$  defined locally by 
$p_r=\frac{\partial L}{\partial {\dot q}^r}$. In the cases
of interest in this paper, the Lagrangian $L$ is singular,
{\it i.e}, the Legendre map 
$\Phi:T{\cal Q}\rightarrow T^*{\cal Q}$ is not onto.
Its image $\Sigma$ is a proper subset of $T^*{\cal Q}$: 
$\Phi(T{\cal Q})=\Sigma\subset T^*{\cal Q}$ and there are 
constraints on the phase space. 
Such situations
are dealt with in Dirac's theory
of constrained
systems \cite{general}.
One iteratively demands preservation of the constraints and this
leads, in general, to more constraints. The algorithm terminates
when no new constraints emerge. The total set of constraints are
divided into first and second class and we suppose that the
second class constraints are eliminated by passage to the Dirac
bracket. An elegant way to do this is to  use the 
Bergmann-Komar starring procedure\cite{BK}.  One simply 
replaces all phase space functions by their starred counterparts.
After the Dirac constraint analysis ends, one has a constrained
Hamiltonian formulation which has the following ingredients:
i) the basic variables (or fields, in a field theory) 
are $(q^r, p_r)$ which span the phase space obeying commutation
relations
\footnote{
these may not be canonical if some 
second class constraints have been eliminated.}. 
ii) a physical interpretation for $q^r$ and $p_r$ that derives
from their definitions as functions of $q$ and ${\dot q}$.
iii) a set of constraints which emerge from 
the constraint analysis.
iv) A Hamiltonian function on the phase space, which
generates the dynamics and preserves the constraints. The Hamiltonian
is arbitrary to the extent of a primary first 
class constraint\footnote{
We do not follow Dirac's suggestion of ``extending'' the Hamiltonian by
adding arbitrary combinations
of the secondary first class constraints to it,
since we wish to stick with
the Lagrangian starting point. This equally means that we do not ``extend''
the symmetry vector field in a similar manner.}.

Let us recapitulate a few known results \cite{symlag,KK,MMS} 
about the 
continuous symmetries of singular Lagrangian systems. Let $S^r(q,{\dot q},t)$
be a symmetry transformation. By this we mean that the change
$\delta_S L$ in the Lagrangian under the changes
$\delta_S q^r=\epsilon S^r(q,{\dot q},t)$, 
$\delta_S {\dot q}^r =\epsilon {\dot S}^r$
in $(q^r,{\dot q}^r)$ is given by a total divergence:
\begin{equation}
\delta_S L=\epsilon \frac{d F(q,{\dot q},t)}{dt}.
\label{symm}
\end{equation}
(Note that in (\ref{symm}) we do not use the 
Euler-Lagrange equations, the accelerations are unrestricted.)
From (\ref{symm}) it follows that on solutions to
the equations of motion, 
the quantity $G_{\cal L}(q,{\dot q},t):=
\frac{\partial L}{\partial {\dot q}^r} S^r-F$ is conserved as a result of N{\"o}ther's theorem.
(\ref{symm}) also implies that $G_{\cal L}(q,{\dot q},t)$ is 
projectable \cite{KK,MMS}
under the Legendre map and therefore can be expressed
as the pull back of a 
function on $\Sigma$  : $G_{\cal L}=\Phi^* G$, . 
In general, the symmetry vector
field $X_S:=S^r\frac{\partial }{\partial q^r} +
{\dot S}^r\frac{\partial }{\partial {\dot q}^r}$ 
(which is defined on $T{\cal Q}$ by using the equations of
motion, or more briefly, the dynamics $\Delta$)
 is {\it not } $\Phi$ projectable\footnote{
The non-projectability of a symmetry transformation has also been recently
remarked  in  \cite{shepley1}.
While these papers too address the question of interplay between gauge 
symmetries and diffeomorphism invariance, their motivation is different from
ours: they seek to find combinations of diffeos and gauge which {\it are}
$\Phi$ projectable.  Our interest here is in {\it pure} diffeo's , which 
in general are not projectable.}. The vertical
 part of $X_S$ projects down to zero, and the horizontal part 
can be expressed 
in the form 
\begin{eqnarray}
\delta_S q^r&=&\epsilon(\frac{\partial G}{\partial p_r}+u^\rho 
\frac{\partial \phi_\rho}{\partial p_r})\nonumber\\
 \delta_S p_r&=&-\epsilon (\frac{\partial G}{\partial q^r}+u^\rho 
\frac{\partial \phi_\rho}{\partial q^r})
\label{delta}
\end{eqnarray}
where $\phi_\rho (q,p)$ are the primary constraints. 
The functions $u^\rho$ are functions on $T{\cal Q}$,
which are not in  general projectable under $\Phi$. 
The non-projectability of $X_S$ has been isolated in the 
functions $u^\rho$, which depend not only on the phase space 
variables $(q^r,p_r)$, but also
the ``unsolved velocities'' $v^\rho$. As the Dirac analysis 
proceeds, the dynamics, and with it
the symmetry vector field (which depends on the dynamics),  
gets
more sharply determined \cite{MMS}. From the 
basic identity (\ref{symm}) it follows that the symmetry 
is ``compatible'' with the dynamics throughout the 
constraint analysis: if the dynamics preserves
constraints, so does the symmetry. The symmetry generator
of $X_S$ is ${\cal G}_S=G+u^\rho \phi_\rho$,
which Kamimura \cite{KK}
refers to as a Generalised Canonical Quantity 
because the $u(q,p,v)$ are not strictly
phase space functions. (They depend on the unsolved velocities $v$).
Symmetries of the Lagrangian translate
into the following properties of the constrained Hamiltonian
formulation, which hold {\it on shell}, (i.e, modulo the equations of motion):
\begin{description}
\item
a) The Lie algebra of the symmetry group is reflected in the
bracket relations of the symmetry generators $\cal G$.

\item
b) The basic variables $q^r$ and $p_r$ are functions on $T{\cal Q}$ and 
transform in a definite manner under the symmetry tranformation
$S$. This transformation property
is reflected
in the bracket relations between these basic variables
and ${\cal G}_{S}$.
\begin{eqnarray}
\delta_S q^r&=&\epsilon\{q^r,{\cal G}_S\}\nonumber\\
\delta_S p_r&=&\epsilon\{p_r,{\cal G}_S\},
\label{delta1}
\end{eqnarray}
\end{description}
where $\{ ,\}$ refers to the Dirac bracket 
resulting from elimination of second class constraints (if any).

In the rest of this paper we apply these general considerations
to the case of interest to us. We consider constrained
Hamiltonian formulations of General Relativity
and the symmetry of interest
is diffeomorphism invariance. 
In this case, as is well known, the generators ${\cal G}_S$ 
are a linear combination of constraints. The criteria listed
above can be used to test for invariance even in a purely Hamiltonian
framework i.e, even when  a Lagrangian is absent. Below, we
will slightly weaken them to allow for the possibility that
they are satisfied modulo gauge\footnote{
In this paper, we reserve the word `gauge' to mean ``internal''
gauge. Diffeomorphisms will not be referred to as `gauge' transformations.
} transformations\footnote{
In this case the diffeomorphism group is twisted with the internal
gauge group. This point is discussed further in the concluding
section.}

\section{Diffeomorphism Invariant Formulations}
We now examine some constrained Hamiltonian formulations
of diffeomorphism invariant theories to see that they 
do indeed satisfy the conditions listed above.
All of these formulations are derived from diffeomorphism 
invariant Lagrangians. Let $({\cal M},g_{\mu\nu}),\mu=0,1,2,3$ 
be a space-time manifold, 
topologically ${\cal S}\times {\mbox{$I\!\!\!\!R$}}$. To 
simplify matters,
we will assume that ${\cal S}$ has no boundary so that we
don't need to keep track of spatial boundary terms.
We are also interested only in infinitesimal diffeomorphisms
and deal entirely with the Lie Algebra rather than the Lie Group
of diffeomorphisms. These infinitesimal diffeomorphisms are generated
by constraints. The constraint algebra ensures that a) is satisfied.
The property a) is discussed extensively \cite{Tei,Ku1,Ku2,Ho,
Ku3,Po,Cla} in the canonical gravity literature as ``path independence''
of evolution and we do not dwell on it any further. We wish
to concentrate on the condition b), which is perhaps implicitly assumed
to be true in the above references. 
 From the general discussion of the 
last section, we expect that the CHF's will satisfy the
conditions a) and b) above. We write down condition b)
explicitly in a few concrete cases and note that it {\it is}
satisfied.

{\it ADM formulation:}
The ADM formulation consists of the following ingredients:
the basic variables are $(q_{ab},\tilde{ \pi}^{ab})$,
which are canonically conjugate. $q_{ab}$ is the pullback
of the space-time metric to a spatial slice ${\cal S}$ and
$\tilde{\pi}^{ab}$ is its conjugate momentum.
The basic variables ($q_{ab},\tilde{\pi}^{ab}$) are subject to the
Hamiltonian constraint
\[
{\tilde {\cal H}} = \frac{1}{{\sqrt{q}}}
(\tilde{\pi}^{ab} {\tilde \pi}_{ab} - \frac{1}{2} 
\tilde{\pi}^2 )-  {{\sqrt{q}}} {}^{3}\!R \approx 0
\]
and the spatial diffeomorphism constraint
\[
\tilde{\cal H}^{b} = {D}_{a} \tilde{\pi}^{ab} \approx 0
\]
where $D$ is the covariant derivative compatible with the
three-metric $q_{ab}$.

The condition (b) holds in the ADM formalism,
 as one would expect
from the general analysis of the last section. The
basic variables of the theory are $(q_{ab}, \tilde{\pi}^{ab})$ and they
have definite space-time meaning: 
$q_{ab}$ is the pull-back of the space-time metric to a spatial slice ${\cal S}$.
By using Hamilton's equations of motion we
see that $\tilde{\pi}^{ab}$ is algebraically related to the extrinsic 
curvature of ${\cal S}$. Since the basic fields 
have a clear space-time meaning, they have a definite
transformation property under space-time diffeomorphisms. 
For example, under an infinitesimal
diffeomorphism generated by a vector field $\xi^a$ 
tangent to ${\cal S}$ ($a=1,2,3$ is a spatial index), 
we expect
\[
\delta q_{ab} ({\rm space-time}) = (D_{a} \xi_{b} + D_{b} \xi_{a}).
\]
If $\xi$ is normal to ${\cal S}$, $\xi^{\mu} = 
N\hat{n}^{\mu}$ we expect
\[
\delta q_{ab} ({\rm space-time}) = {\cal L}_{\xi} q_{ab} = N K_{ab},
\]
where $K_{ab}$ is the extrinsic curvature of ${\cal S}$.

One can also compute the change in the basic variables by taking
their Poisson brackets with the 
diffeomorphism generator $C(\xi)$:
\begin{eqnarray}
\delta_{\xi} q_{ab} ({\rm canonical}) &=& \{ q_{ab},
 C(\xi) \},\nonumber\\
\delta_{\xi} \tilde{\pi}^{ab} ({\rm canonical}) &=& \{ \tilde{\pi}^{ab},
 C(\xi) \}.
\end{eqnarray}
The condition (b) is satisfied in the ADM formalism 
since \cite{carlip} as 
follows from Hamilton's equations
\begin{eqnarray*}
\delta_{\xi} q_{ab} ({\rm space-time}) &=& \delta_{\xi} q_{ab} ({\rm canonical})\\
\delta_{\xi} \tilde{\pi}^{ab} ({\rm space-time}) &=& \delta_{\xi} {\tilde \pi}^{ab} ({\rm canonical}).
\end{eqnarray*}

{\it 2+1 Palatini gravity:}
The next example we consider is gravity in 2+1 
dimensions in its Palatini formulation. The basic fields are
$e_\mu^I$ and $A_\mu^{IJ}$; $\mu=0,1,2$ is a tangent space
index and $I = 0,1,2$ is an internal Minkowski index. 
$e_\mu^I$ is a triad and $A_\mu^{IJ}$
an $SO(2,1)$ connection.
The action  is given by 
\[
I = \frac{1}{2} \int e_{I} \wedge F^{I},
\]
where $F=dA+A\wedge A$ in the 
notation of differential forms. A standard
constraint analysis leads to the following Hamiltonian 
formulation: The basic variables are the canonically 
conjugate pair $({\tilde e}^{I}{}^{a}:={\tilde 
\eta}^{ab} e_b^I,A^{I}_{a})$, where
$a$ is a spatial index. The constraints of the theory are
\begin{eqnarray*}
F^{I} &=& 0\\
G^{I} &=& {\cal D} \wedge e^{I} = 0,
\end{eqnarray*}
where it is understood that these two-forms are pulled back 
to a spatial slice ${\cal S}$.
Diffeomorphism are generated by combinations of constraints. If 
$\xi^{\mu}$ is a vector field on ${\cal M}$,
\[
C(\xi) = \int_{\cal S}(\xi^{\mu} e^{I}_{\mu} F_{I} + \xi^{\mu} A^{I}_{\mu} G_{I})
\]
generates a pure diffeomorphism on the basic variables.
It is easily checked, using the $ISO(2,1)$ algebra satisfied by the
constraints that the condition (b) above is satisfied
on shell (using the equations of motion).

{\it Extended phase space construction (EPS)}: As a last example, we consider
the extended phase space of Ashtekar. This CHF was originally arrived
at by Ashtekar \cite{Ashtate} by extending the ADM phase space to 
incorporate triads. This example is instructive because it can also be 
derived \cite{Ashtate,JO,Joe} from a 
manifestly covariant Lagrangian {\it by fixing the ``time gauge"}. This
example will illustrate how internal gauge fixing interacts with diffeomorphism
invariance. As we will
see, because of the gauge fixing a) and b) are not satisfied as they stand
but they {\it are} satisfied   
modulo $SO(3)$ gauge.

Let us start with the following action principle. The basic fields are 
$e^{I}_{\mu},\;\;A^{IJ}_{\mu}$, where $e^{I}_{\mu}$ is a tetrad field
and $A^{IJ}_{\mu}$ an $SO(3,1)$ connection field. The action is
\begin{equation}
I = \frac{1}{2} \int e^{I} \wedge e^{J} \wedge F^{KL} \epsilon_{IJKL},
\end{equation}
where we use differential form notation and $F = dA + A\wedge A$. A 
straight forward Legendre transformation \cite{Ashtate} results in the 
following CHF. The basic conjugate variables are $(A_a, \tilde{\alpha}^{a})$ 
where
\begin{equation}
\tilde{\alpha}^{a}_{IJ} = \tilde{\eta}^{abc} e_{bI} e_{cJ}.
\end{equation}
These variables are subject to the constraints
\begin{equation}
G_{IJ} = {\cal D}_{a} \tilde{\alpha}^{a}_{IJ} \approx 0 \label{Gauss}
\end{equation}
\begin{equation}
V_{a} = T r \tilde{\alpha}^{b} F_{ab} \approx 0 \label{vector}
\end{equation}
\begin{equation}
S = T r \tilde{\alpha}^{a} \tilde{\alpha}^{b} F_{ab} \approx 0 \label{scalar}
\end{equation}
\begin{equation}
\phi^{ab}: = \epsilon^{IJKL} \tilde{\alpha}_{IJ}^{a} \tilde{\alpha}^{b}_{KL} \approx 0 \label{simple}
\end{equation}
\begin{equation}
\chi^{ab}: = \epsilon^{IJKL} \tilde{\alpha}^{cM}_{I} \tilde{\alpha}^{(a}_{MJ} ({\cal D}_{c} \tilde{\alpha}^{b)})_{KL} \approx 0. \label{rec}
\end{equation}
Of these the last two (\ref{simple}, \ref{rec}) are second class.
Let us suppose these second class constraints to be formally
eliminated by passing to the Dirac bracket. No gauge fixing has been
done so far and it follows from the general theory 
summarised in the last section that the Hamiltonian 
formulation above satisfies a) as well as b).

(\ref{simple}) implies \cite{Ashtate} that ${\tilde \alpha}_{IJ}$ is
of the form
${\tilde E}^a{}_{[I} n_{J]}$ for {\it some} internal vector $n_J$. 
Let us now impose the ``time" gauge, i.e., pick $n_{I}$ to have the 
standard form ${\stackrel{\circ}{n}}^{I} = (1,0,0,0)$. This corresponds to
choosing $e^0$ to be normal to the spatial slice ${\cal S}$.
One is, of course, at liberty to make this gauge choice. In order
to enforce this gauge choice, we need to impose a constraint
\begin{equation}
\chi^{I} = n^{I} - {\stackrel{\circ}{n}}^{I}\approx 0. \label{time}
\end{equation}
This constraint breaks the $SO(3,1)$ gauge generated by the 
 Gauss law constraint (\ref{Gauss}) down to $SO(3)$. The ``Boost part" 
\begin{equation}
B_{I} = G_{IJ} {\stackrel{\circ}{n}}^{J} \label{boost}
\end{equation}
of (\ref{Gauss}) does not commute with (\ref{time}) and in fact 
$(B_{I}, \chi^{I})$ form a second class set. If one eliminates this second 
class set one arrives at EPS. Writing $i,j$ instead of $I,J$
for indices orthogonal to ${\stackrel{\circ}{n}}^I$,
we find that basic variables of EPS are
$({\tilde E}^a{}_i,K_a{}^i)$ which are canonically conjugate
and have the
space-time interpretation of densitised 
triad and extrinsic curvature respectively. The constraints of
the theory are:

\begin{eqnarray*}
\epsilon_{ijk} K^{j}_{a} \tilde{E}^{ak} &\approx& 0 \label{SO(3)}\\
D_{a} [\tilde{E}^{a}_{k} K^{k}_{b} - \delta^{a}_{b} \tilde{E}^{c}_{k}
K^{k}_{c}] &\approx& 0\\
{\sqrt{q}} R + \frac{2}{\sqrt{q}} \tilde{E}^{[a}_{i}
\tilde{E}^{b]}_{j} K^{i}_{a} K^{j}_{b} &\approx& 0,
\end{eqnarray*}
where $D_{a}$ is the covariant derivative associated with
$q_{ab}$ and $R$, its scalar curvature.

Are conditions a) and b) satisfied in the gauge fixed theory?
Diffeomorphisms that displace ${\cal S}$
normal to itself will in general, spoil the ``time gauge". In order to 
restore the ``time gauge" (and this is the BK starring procedure
of passing to Dirac brackets) one has to add some definite linear combination 
of $B_{I}$ to the diffeomorphism generator. As a result, 
(since the commutator of two boosts is a rotation) the 
diffeomorphism algebra closes only up to $SO(3)$ gauge rotations. In
the same way, (b) is only satisfied up to $SO(3)$ gauge rotations. We 
describe this theory as satisfying a) and b) (mod $SO(3)$ gauge). 
The lesson to be
learned from this example is that if one derives a 
Hamiltonian formulation from a Lagrangian and fixes gauge 
in the derivation, the resulting Hamiltonian formulation is diffeomorphism
invariant(modulo gauge).

\section{Strong and Weak Diffeomorphism Invariance}

It is clear from these examples that diffeomorphism invariance in the 
Hamiltonian framework means {\it more} than getting the constraint
algebra right.
It is also necessary that under the 
action of the diffeomorphism generators, the basic variables must
transform as expected from their space-time interpretation. We will 
refer to a CHF which satisfies the first condition (a) as weakly
diffeomorphism invariant. A theory that also satisfies (b) is called
strongly diffeomorphism invariant. It is clear that before we can test
a CHF for diffeomorphism invariance, the space-time meaning of the basic
variables has to be declared, since condition (b) explicitly needs this
knowledge.

 To better understand the meaning of Strong Diffeomorphism
 invariance, it is useful to consider a simpler
 but analogous situation: classical 
 relativistic particle dynamics \cite{Dirac,Komar,Todorov}. Direct
 interactions between $N$ relativistic
 particles in Minkowski space can be described by mathematical
 models 
 which are constrained Hamiltonian formulations.
  The models are defined as follows: the basic variables
  are $(x_a{}^\mu,p_{a\mu}), (a=1..N, \mu=0,1,2,3)$, where
  $a$ is particle index (for the duration of this section)
  and $\mu$ a Minkowski space-time
  index. One can define the system by imposing $2N$ second class
  constraints. 
  The constraints are needed to reduce the phase space
  degrees of freedom from $8N$ to $6N$, which is the right
  number for $N$ particles. The symmetry of interest here
  is the Poincare group. We will say that a model
  is Poincare invariant if the following conditions hold:
  \begin{description}
  \item a)There exist $10$ functions (one for each of the 
  Poincare generators) on the phase space whose Dirac brackets
  reflect the Lie Algebra of the Poincare group.
  
  \item b) The Dirac brackets between the basic variables
   $(x^\mu{}_a,p_{a\mu})$ and the Poincare generators reflect the
   space-time transformation properties of the basic variables.
  \end{description}
As was first pointed out by Pryce \cite{Pryce}, Poincare
invariance means {\it both} a) and b) and not just a). Bakamjian
and Thomas\cite{BT} were able to construct
interacting models, but at the cost of giving up
condition b). In these models \cite{BT},
particle world lines would depend on the Lorentz frame of the observer.
(To clarify this point, it is not just the same world line viewed
from different Lorentz frames, but {\it different} world lines.) 
 This amounts to giving up the objectivity of world lines, or
 particle histories, which is unacceptable, since classically, 
 particle world lines can be experimentally measured 
\footnote{
There are models (see model 1 of this paper) 
in which these conditions are satisfied modulo
reparametrisation gauge. This is quite acceptable since, it
does not compromise the objectivity of particle World lines. All that
happens is that the World line is reparametrised under a 
Poincare transformation.}.

To clarify the meaning of the conditions a) and b) above, we discuss two
models  which are in the literature \cite{Komar,Todorov,SMG,Samuel}. 
Both models describe two interacting particles and are defined
by imposing $4$ constraints on the sixteen dimensional phase space
spanned by $(x_a{}^\mu, p_a{}_\mu)$, the position and momentum four
vectors of the particles.

{\it model 1} The constraints that define this model are
\begin{eqnarray}
K_1&=&p_1^2+m_1^2+V((x_1-x_2)^2)\\
K_2&=&p_2^2+m_2^2+V((x_1-x_2)^2)\\
\chi_1&=& P.(x_1-x_2)\\
\chi_2&=& P.x_1-\tau,\\
\end{eqnarray}
 where $P=p_1+p_2$ is the total momentum four-vector of the two 
 particles, $V((x_1-x_2)^2)$ is a potential function which depends only on
 the invariant interval $(x_1-x_2)^2$ between the two particle position four vectors and $\tau$
 is an evolution parameter, which plays the role of ``time''.
 The constraints reduce the dimension of the phase space to 12
 and do this in a Poincare invariant manner: both 
 a) and b) above are satisfied. To see this, note that $P_\mu$ 
 and $M_{\mu\nu}=\Sigma_{a=1,2}(x_a{}_\mu p_a{}_\nu-x_a{}_\nu p_a{}_\mu)$,
 the ten generators of the Poincare group commute with
 $K_1,K_2$ and $\chi_1$ (with all but one of the constraints). It
 follows from this (and the definition of the Dirac bracket) that 
 a) above is satisfied. It also follows that b) is satisfied modulo
 reparametrisation, since the Dirac bracket of the basic variables
 $(x_a{}^\mu, p_a{}_\mu)$ with the Poincare generators agrees with the
 Poisson bracket (which in turn agrees with the four-vector
  space-time transformation
 property of $(x_a{}^\mu, p_a{}_\mu)$) apart 
 from a term representing the reparametrisation of the world line.
 Writing $G$ for any one of the ten generators of the Poincare group,
 \begin{equation}
 \{x_a{}^\mu,G\}^*= \{x_a{}^\mu,G\}+\frac{dx_a{}^\mu}{d\tau} \delta_a \tau
 \label{WLC}
 \end{equation}
 for some $\delta_a \tau$.
 
 {\it model 2} The second model is defined by the constraints
 \begin{eqnarray}
K_1&=&p_1^2+m_1^2+V(x_1,x_2,p_1,p_2)\\
K_2&=&p_2^2+m_2^2+V(x_1,x_2,p_1,p_2)\\
\chi_1&=& (x_1-x_2)^0\\
\chi_2&=& (x_1)^0-\tau,\\
\end{eqnarray}
 $K_1$ and $K_2$ are required to commute with each other 
 \cite{Komar,Todorov,SMG} and with
 all the Poincare generators. From the definition of the Dirac bracket
 it follows that this model satisfies a). However, it does {\it not }
 satisfy b). It must therefore be rejected as a description
 of two relativistic particles.
 
The analogy between the models described above and canonical gravity
is as follows: The symmetry group of interest in the first case is the
Poincare group and in the second case the diffeomorphism group. The classical
histories of the first system describe the world lines of N particles and
in the second case a space--time. In both cases, the problem is one of 
realising a symmetry group in a purely Hamiltonian framework.
Our main point here is that it is possible to get the symmetry algebra right
but still violate the symmetry by giving up b). Model 2 is an example of this.

Returning to our problem in canonical gravity, a CHF which
is only weakly diffeomorphism invariant suffers from the following
feature:
Given a space-time history (a solution of the field equations), one can 
slice it up in many ways in a 3+1 formalism. Conversely, 
given initial data and  particular
slicing one can evolve the initial data and produce a ``history''
by ``stacking'' the spatial slices in temporal order\cite{MTW}. In theories
where b) is given up the ``history'' 
which is produced {\it depends
on the slicing}. This means that the history has no 
objective reality. One would of course like measureable quantities
to have an objective meaning (independent of slicing). One should
therefore be aware of which fields in a theory are objectively real. For example,
in the EPS, the fields $q_{ab}$ and $K_{ab}$ (which are $SO(3)$ gauge 
invariant) {\it do} have objective reality. But the basic fields
in the formulation $({\tilde E}^a{}_i,K_a{}^i)$ do not. They are only defined 
modulo $SO(3)$ gauge.

\section{Conclusion}
We have shown that for a constrained Hamiltonian
formulation of gravity to be diffeomorphism invariant
there must be diffeomorphism generators
on the phase space  so that 
a) the generators reflect the 
algebra of the diffeomorphism group
in their brackets and b) the space-time interpretation
of the basic fields is reflected in their brackets with
the diffeomorphism generators.
These conditions are automatically
satisfied by CHFs which derive from a covariant Lagrangian. In the 
absence of a Lagrangian these conditions can be used to test for diffeomorphism
invariance. Condition a) has been emphasised in the literature, but it
appears that condition b) is usually left implicit. In this paper we 
point out what goes wrong if one gives up condtion b): one loses the 
objectivity of history.

Notice that a covariant Lagrangian automatically gives us space-time
interpretations for all the  phase space variables appearing in
the Hamiltonian formulation. In a purely Hamiltonian approach
one has to not only prescribe the basic variables, their brackets,
and the constraints, but also give a space-time interpretation for the
basic variables. Unless this is done, it is not possible to
physically interpret the Hamiltonian system. If the PB of the 
diffeomorphism generator with a phase space variable
does not reflect its space-time interpretation, one loses
a space-time interpretation for that variable even at the classical
level.

The Diffeomorphism 
invariance of the theory can only be decided after the space-time interpretation
of the basic variables has been declared (since condition b) explicitly
needs this knowledge).  Indeed, unless the space-time interpretation of the
basic variables is declared, the CHF is not even fully defined.
A CHF may be diffeomorphism invariant
with one interpretation and not invariant with another space-time
interpretation of the basic variables. An example of this phenomenon
is discussed in \cite{ham}. Barbero's Hamiltonian formulation of General
Relativity {\it is} strongly diffeomorphism invariant with another 
space--time interpretation of the basic variables, but not with 
the space--time ``gauge field interpretation'' that one might 
prefer. In contrast, Ashtekar's Hamiltonian formulation is SDI 
with both interpretations for the connection variable: that deriving
from the canonical transformation as well as for the space--time
gauge field interpretation.

In the EPS model, the Lie Algebra of the Diffeomorphism group is not
a subalgebra of the constraint generators, but appears as a quotient.
We arrived at conditions a) and b) by assuming that the symmetry
group of interest was a {\it subgroup} of the total Lagrangian symmetry
group. The known Lagrangian formulations of  General Relativity 
all have the property that the diffeomorphism group is a subgroup. One
can slightly relax this assumption and allow for ``twisted products'',
where the diffeomorphism group only appears as a quotient. The situation
then is very similar to the EPS formulation, where the diffeomorphism
Lie Algebra only closes modulo gauge.

One may object that one should not demand that the basic variables
be objectively defined in space-time, since they are not ``observables''
in the Dirac sense. This objection is easily met:  it is easy to
construct ``observables'' from the basic variables by using a device 
explained in \cite{Gaul}. Although $q_{ab}$ is not an ``observable'', 
the distance between invariantly specified events is an ``observable''.
E.g, one can locate an event as the intersection of two particle 
world lines or (in the absence of matter) as a point where
four scalars constructed from the gravitational field 
\cite{Berg}
vanish. If a CHF is strongly diffeomorphism invariant in the sense of this paper,
such ``observables'' do have an objective meaning. Otherwise, the answer
predicted by the CHF could depend on slicing. A CHF which
violates strong diffeomorphism invariance classically 
should be rejected as an unsuitable starting point for building
a quantum theory.

{\it Acknowledgement:} It is a pleasure to thank Richard Epp,
B.R. Iyer, Sukanya Sinha and Madhavan Varadarajan for 
extended discussions. 

\end{document}